%% file: quote-eval.tex
\documentclass[fleqn]{llncs}

\usepackage{latexsym}
\usepackage{amssymb}
\usepackage{stmaryrd}
\usepackage{phonetic}
\usepackage{wasysym}
\usepackage{pgf}
\usepackage{tikz}
\usepackage{url}
\usetikzlibrary{arrows}

\input{llncs-def}

\title{The Formalization of Syntax-Based Mathematical Algorithms Using
  Quotation and Evaluation\thanks{Appears in \emph{Intelligent
      Computer Mathematics} (proceedings of CICM 2013), \emph{Lecture
      Notes in Computer Science}, Vol.~7961, pp.~35--50,
    Springer-Verlag, 2013.  The final publication is available at
    \texttt{link.springer.com}.  This research was supported by
    NSERC.}}

\author{William M. Farmer\thanks{\texttt{wmfarmer@mcmaster.ca}.}}

\institute{%
Department of Computing and Software\\
McMaster University\\
Hamilton, Ontario, Canada
\\[1.5ex]
5 August 2013
}

\begin{document} 
\maketitle

\begin{abstract}
Algorithms like those for differentiating functional expressions
manipulate the syntactic structure of mathematical expressions in a
mathematically meaningful way.  A formalization of such an algorithm
should include a specification of its computational behavior, a
specification of its mathematical meaning, and a mechanism for
applying the algorithm to actual expressions.  Achieving these goals
requires the ability to integrate reasoning about the syntax of the
expressions with reasoning about what the expressions mean.  A
\emph{syntax framework} is a mathematical structure that is an
abstract model for a syntax reasoning system.  It contains a mapping
of expressions to \emph{syntactic values} that represent the syntactic
structures of the expressions; a language for reasoning about
syntactic values; a \emph{quotation} mechanism to refer to the
syntactic value of an expression; and an \emph{evaluation} mechanism
to refer to the value of the expression represented by a syntactic
value.  We present and compare two approaches, based on instances of a
syntax framework, to formalize a syntax-based mathematical algorithm
in a formal theory $T$.  In the first approach the syntactic values
for the expressions manipulated by the algorithm are members of an
inductive type in $T$, but quotation and evaluation are functions
defined in the metatheory of $T$.  In the second approach every
expression in $T$ is represented by a syntactic value, and quotation
and evaluation are operators in $T$ itself.
\end{abstract}

\section{Introduction}

A great many of the algorithms employed in mathematics work by
manipulating the syntactic structure of mathematical expressions in a
mathematically meaningful way.  Here are some examples:

\be

  \item Arithmetic operations applied to numerals.

  \item Operations such as factorization applied to polynomials.

  \item Simplification of algebraic expressions.

  \item Operations such as transposition performed on matrices.

  \item Symbolic differentiation and antidifferentiation of
    expressions with variables.

\ee
The study and application of these kinds of algorithms is called
\emph{symbolic computation}.  For centuries symbolic computation was
performed almost entirely using pencil and paper (and similar
devices).  However, today symbolic computation can be performed by
computer, and algorithms that manipulate mathematical expressions are
the main fare of computer algebra systems.

In this paper we are interested in the problem of how to formalize
syntax-based mathematical algorithms.  These algorithms manipulate
members of a formal language in a computer algebra system, but their
behavior and meaning are usually not formally expressed in a computer
algebra system.  However, we want to use these algorithms in formal
theories and formally understand what they do.  We are interested in
employing existing external implementations of these algorithms in
formal theories as well as implementing these algorithms directly in
formal theories.

As an illustration, consider an algorithm, say named \mname{RatPlus},
that adds rational number numerals, which are represented in memory in
some suitable way.  (An important issue, that we will not address, is
how the numerals are represented to optimize the efficiency of
\mname{RatPlus}.)  For example, if the numerals $\frac{2}{5}$ and
$\frac{3}{8}$ are given to \mname{RatPlus} as input, the numeral
$\frac{31}{40}$ is returned by \mname{RatPlus} as output.  What would
we need to do to use \mname{RatPlus} to add rational numbers in a
formal theory $T$ and be confident that the results are correct?
First, we would have to introduce values in $T$ to represent rational
number numerals as syntactic structures, and then define a binary
operator $O$ over these values that has the same input-output relation
as \mname{RatPlus}.  Second, we would have to prove in $T$ that, if
$O(a,b) = c$, then the sum of the rational numbers represented by $a$
and $b$ is the rational number represented by $c$.  And third, we
would have to devise a mechanism for using the definition of $O$ to
add rational numbers in $T$.

The second task is the most challenging.  The operator $O$, like
\mname{RatPlus}, manipulates numerals as syntactic structures.  To
state and then prove that these manipulations are mathematically
meaningful requires the ability to express the interplay of how the
numerals are manipulated and what the manipulations mean with respect
to rational numbers.  This is a formidable task in a traditional logic
in which there is no mechanism for directly referring to the syntax of
the expressions in the logic.  We need to reason about a rational
number numeral $\frac{2}{5}$ both as a syntactic structure that can be
deconstructed into the integer numerals 2 and 5 and as an expression
that denotes the rational number $2/5$.

Let us try to make the problem of how to formalize syntax-based
mathematical algorithms like \mname{RatPlus} more precise.  Let $T$ be
a theory in a traditional logic like first-order logic or simple type
theory, and let $A$ be an algorithm that manipulates certain
expressions of $T$.  To formalize $A$ in $T$ we need to do three
things:

\be

  \item \emph{Define an operator $O_A$ in $T$ that represents $A$}:
    Introduce values in $T$ that represent the expressions manipulated
    by $A$.  Introduce an operator $O_A$ in $T$ that maps the values
    that represent the input expressions taken by $A$ to the values
    that represent the output expressions produced by $A$.  Write a
    sentence named \mname{CompBehavior} in $T$ that specifies the
    \emph{computational behavior} of $O_A$ to be the same as that of
    $A$.  That is, if $A$ takes an input expression $e$ and produces
    an output expression $e'$, then \mname{CompBehavior} should say
    that $O_A$ maps the value that represents $e$ to the value that
    represents $e'$.

  \item \emph{Prove in $T$ that $O_A$ is mathematically correct}:
    Write a sentence named \mname{MathMeaning} in $T$ that specifies
    the \emph{mathematical meaning} of $O_A$ to be the same as that of
    $A$.  That is, if the value of an input expression $e$ given to
    $A$ is related to the value of the corresponding output expression
    $e'$ produced by $A$ in a particular way, then \mname{MathMeaning}
    should say that the value of the expression representing $e$
    should be related to the value of the expression representing $e'$
    in the same way.  Finally, prove \mname{MathMeaning} from
    \mname{CompBehavior} in $T$.

  \item \emph{Devise a mechanism for using $O_A$ in $T$}: An
    application $O_A(a_1,\ldots,a_n)$ of $O_A$ to the values
    $a_1,\ldots,a_n$ can be used in $T$ by instantiating
    \mname{MathMeaning} with $a_1,\ldots,a_n$ and then simplifying the
    resulting formula to obtain a statement about the value of
    $O_A(a_1,\ldots,a_n)$.  For the sake of convenience or efficiency,
    we might want to use $A$ itself to compute $O_A(a_1,\ldots,a_n)$.
    We will know that results produced by $A$ are correct provided $A$
    and $O_A$ have the same computational behavior.

\ee

If we believe that $A$ works correctly and we are happy to do our
computations with $A$ outside of $T$, we can skip the writing of
\mname{CompBehavior} and use \mname{MathMeaning} as an axiom that
asserts $A$ has the mathematical meaning specified by
\mname{MathMeaning} for $O_A$.  The idea of treating specifications of
external algorithms as axioms is a key component of the notion of a
\emph{biform theory}~\cite{CaretteFarmer11,Farmer07b}.

So to use $A$ in $T$ we need to formalize $A$ in $T$, and to do this,
we need a system that integrates reasoning about the syntax of the
expressions with reasoning about what the expressions mean.  A
\emph{syntax framework}~\cite{FarmerLarjani13} is a mathematical
structure that is an abstract model for a syntax reasoning system. It
contains a mapping of expressions to \emph{syntactic values} that
represent the syntactic structures of the expressions; a language for
reasoning about syntactic values; a quotation mechanism to refer to
the syntactic value of an expression; and an evaluation mechanism to
refer to the value of the expression represented by a syntactic value.
A syntax framework provides the tools needed to reason about the
interplay of syntax and semantics.  It is just what we need to
formalize syntax-based mathematical algorithms.

\emph{Reflection} is a technique to embed reasoning about a reasoning
system (i.e., metareasoning) in the reasoning system itself.
Reflection has been employed in logic~\cite{Koellner09}, theorem
proving~\cite{Harrison95}, and programming~\cite{DemersMalenfant95}.
Since metareasoning very often involves the syntactic manipulation of
expressions, a syntax framework is a natural subcomponent of a
reflection mechanism.

This paper attacks the problem of formalizing a syntax-based
mathematical algorithm $A$ in a formal theory $T$ using syntax
frameworks.  Two approaches are presented and compared.  The first
approach is local in nature.  It employs a syntax framework in which
there are syntactic values only for the expressions manipulated by
$A$.  The second approach is global in nature.  It employs a syntax
framework in which there are syntactic values for all the expressions
of $T$.  We will see that these two approaches have contrasting
strengths and weaknesses.  The local approach offers an incomplete
solution at a low cost, while the global approach offers a complete
solution at a high cost.

The two approaches will be illustrated using the familiar example of
polynomial differentiation.  In particular, we will discuss how the
two approaches can be employed to formalize an algorithm that
differentiates expressions with variables that denote real-valued
polynomial functions.  We will show that algorithms like
differentiation that manipulate expressions with variables are more
challenging to formalize than algorithms like symbolic arithmetic that
manipulate numerals without variables.

The following is the outline of the paper.  The next section,
Section~\ref{sec:diff}, presents the paper's principal example,
polynomial differentiation.  The notion of a syntax framework is
defined in Section~\ref{sec:syntax}.  Sections~\ref{sec:local}
and~\ref{sec:global} present the local and global approaches to
formalizing syntax-based mathematical algorithms.  And the paper
concludes with Section~\ref{sec:conclusion}.

\section{Example: Polynomial Differentiation} \label{sec:diff}

We examine in this section the problem of how to formalize a symbolic
differentiation algorithm and then prove that the algorithm actually
computes derivatives.  We start by defining what a derivative is.

Let $f : \mathbb{R} \tarrow \mathbb{R}$ be a function over the real
numbers and $a \in \mathbb{R}$.  The \emph{derivative of $f$ at $a$},
written $\mname{deriv}(f,a)$, is \[\lim_{h \tarrow 0}\frac{f(a + h) -
  f(a)}{h}\] if this limit exists.  The \emph{derivative of $f$},
written $\mname{deriv}(f)$, is the function \[\LambdaApp x \mcolon
\mathbb{R} \mdot \mname{deriv}(f,x).\] Notice that we are using the
traditional definition of a derivative in which a derivative of a
function is defined pointwise.

\emph{Differentiation} is in general the process of finding
derivatives which ultimately reduces to finding limits.
\emph{Symbolic differentiation} is the process of mechanically
transforming an expression with variables that represents a function
over the real numbers into an expression with variables that
represents the derivative of the function.  For example, the result of
symbolically differentiating the expression $\mname{sin}(x^2)$ which
represents the function \mbox{$\LambdaApp x \mcolon \mathbb{R} \mdot
  \mname{sin}(x^2)$} is the expression $2 \cdot x \cdot
\mname{cos}(x^2)$ which represents the function $\LambdaApp x \mcolon
\mathbb{R} \mdot 2 \cdot x \cdot \mname{cos}(x^2)$.  Symbolic
differentiation is performed by applying certain \emph{differentiation
  rules} and \emph{simplification rules} to a starting expression
until no rule is applicable.

Let us look at how symbolic differentiation works on polynomials.  A
\emph{polynomial} is an expression constructed from real-valued
constants and variables by applying addition, subtraction,
multiplication, and natural number exponentiation.  For
example, $x\cdot(x^2 + y)$ is a polynomial.  The symbolic
differentiation of polynomials is performed using the following
well-known differentiation rules:

\bi

  \item[]\textbf{Constant Rule}\[\frac{d}{dx}(c) = 0 \hspace{3ex}
    \mbox{where $c$ is a constant or a variable different from $x$}.\]

  \item[]\textbf{Variable Rule}\[\frac{d}{dx}(x) = 1.\]

  \item[]\textbf{Sum and Difference Rule}\[\frac{d}{dx}(u \pm v) =
    \frac{d}{dx}(u) \pm \frac{d}{dx}(v).\]

  \item[]\textbf{Product Rule}\[\frac{d}{dx}(u \cdot v) =
    \frac{d}{dx}(u) \cdot v + u \cdot \frac{d}{dx}(v).\]

  \item[]\textbf{Power Rule}\[\frac{d}{dx}(u^n) = 
    \left\{\begin{array}{ll}
              0 & 
              \hspace{1ex} \mbox{if } n = 0\\
              n \cdot u^{n-1} \cdot \frac{d}{dx}(u) & 
              \hspace{1ex} \mbox{if } n > 0.  
           \end{array}
    \right.\]

\ei
Written using traditional Leibniz notation, the rules specify how
symbolic differentiation is performed with respect to the variable
$x$.  The symbols $u$ and $v$ range over polynomials that may contain
$x$ as well as other variables, and the symbol $n$ ranges over natural
numbers.  Notice that these rules are not meaning preserving in the
usual way; for example, the rule $\frac{d}{dx}(c) = 0$ is not meaning
preserving if we view $c$ as a value and not as an expression.

Let \mname{PolyDiff} be the algorithm that, given a polynomial $u$ and
variable $x$, applies the five differentiation rules above to the
starting expression $\frac{d}{dx}(u)$ until there are no longer any
expressions starting with $\frac{d}{dx}$ and then simplifies the
resulting expression using the rules $0 + u = u + 0 = 0$ and $1 \cdot
u = u \cdot 1 = u$ and collecting like terms.  Applied to $x \cdot
(x^2 + y)$, \mname{PolyDiff} would perform the following steps:
\begin{eqnarray}
  \frac{d}{dx}(x \cdot (x^2 + y))
  & = & \frac{d}{dx}(x) \cdot (x^2 + y) +
        x \cdot \frac{d}{dx}(x^2 + y)\\
  & = & 1 \cdot (x^2 + y) + 
        x \cdot \Big(\frac{d}{dx}(x^2) + \frac{d}{dx}(y)\Big)\\
  & = & 1 \cdot (x^2 + y) + 
        x \cdot \Big(2 \cdot x^1 \cdot \frac{d}{dx}(x) + 0\Big)\\
  & = & 1 \cdot (x^2 + y) + 
        x \cdot (2 \cdot x^1 \cdot 1 + 0)\\
  & = & 3 \cdot x^2 + y
\end{eqnarray}
Line (1) is by the Product Rule; (2) is by the Variable and Sum and
Difference Rules; (3) is by the Power and Constant Rules; (4) is by
the Variable Rule; and (5) is by the simplification rules.  Thus,
given the function \[f = \LambdaApp x \mcolon \mathbb{R} \mdot x \cdot
(x^2 + y),\] using \mname{PolyDiff} we are able to obtain the
derivative \[\LambdaApp x \mcolon \mathbb{R} \mdot 3 \cdot x^2 + y\]
of $f$ via mechanical manipulation of the expression $x \cdot (x^2 +
y)$.

Algorithms similar to \mname{PolyDiff} are commonly employed in
informal mathematics.  In fact, they are learned and applied by every
calculus student.  They should be as available and useful in formal
mathematics as they are in informal mathematics.  We thus need to
formalize them as described in the Introduction.

The main objective of this paper is to show how syntax-based
mathematical algorithms can be formalized using \mname{PolyDiff} as an
example.  We will begin by making the task of formalizing
\mname{PolyDiff} precise.

Let a \emph{theory} be a pair $T = (L,\Gamma)$ where $L$ is a formal
language and $\Gamma$ is a set of sentences in $L$ that serve as the
axioms of the theory.  Define $T_R = (L_R,\Gamma_R)$ to be a theory of
the real numbers in (many-sorted) simple type theory.  We assume that
$L_R$ is a set of expressions over a signature that includes a type
$\mathbb{R}$ of the real numbers, constants for each natural number,
and constants for addition, subtraction, multiplication, natural
number exponentiation, and the unary and binary \mname{deriv}
operators defined above.  We assume that $\Gamma_R$ contains the
axioms of a complete ordered field as well as the definitions of all
the defined constants in $L_R$ (see~\cite{Farmer08} for further
details).


Let $L_{\rm var} \subseteq L_R$ be the set of variables of type
$\mathbb{R}$ and $L_{\rm poly} \subseteq L_R$ be the set of
expressions constructed from members of $L_{\rm var}$, constants of
type $\mathbb{R}$, addition, subtraction, multiplication, and natural
number exponentiation.  Finally, assume that $\mname{PolyDiff}: L_{\rm
  poly} \times L_{\rm var} \tarrow L_{\rm poly}$ is the algorithm
described in the previous section adapted to operate on expressions of
$L_R$.

Thus to formalize $\mname{PolyDiff}$ we need to:

\be

  \item Define an operator $O_{\rm pd}$ in $T_R$ that represents
    $\mname{PolyDiff}$.

  \item Prove in $T_R$ that $O_{\rm pd}$ is mathematically correct.

  \item Devise a mechanism for using $O_{\rm pd}$ in $T_R$.

\ee

Formalizing $\mname{PolyDiff}$ should be much easier than formalizing
differentiation algorithms for larger sets of expressions that
include, for example, rational expressions and transcendental
functions.  Polynomial functions are total (i.e., they are defined at
all points on the real line) and their derivatives are also total.  As
a result, issues of undefinedness do not arise when specifying the
mathematical meaning of $\mname{PolyDiff}$.  

However, functions more general than polynomial functions as well as
their derivatives may be undefined at some points.  Thus using a
differentiation algorithm to compute the derivative of one of these
more general functions requires care in determining the precise domain
of the derivative.  For example, differentiating the rational
expression $x/x$ using the well-known Quotient Rule yields the
expression 0, but the derivative of $\LambdaApp x \mcolon \mathbb{R}
\mdot x/x$ is not $\LambdaApp x \mcolon \mathbb{R} \mdot 0$.  The
derivative is actually the partial function \[\LambdaApp x \mcolon
\mathbb{R} \mdot \mname{if } x \not= 0 \mname{ then } 0 \mname{ else }
\bot.\] We restrict our attention to differentiating polynomial
functions so that we can focus on reasoning about syntax without being
concerned about issues of undefinedness.

\section{Syntax Frameworks} \label{sec:syntax}

A syntax framework~\cite{FarmerLarjani13} is a mathematical structure
that is intended to be an abstract model of a system for reasoning
about the syntax of an interpreted language (i.e., a formal language
with a semantics).  It will take several definitions
from~\cite{FarmerLarjani13} to present this structure.

\begin{definition}[Interpreted Language] \label{df:interp-lang} \em 
\noindent
An \emph{interpreted language} is a triple $I=(L,D_{\rm sem},V_{\rm
  sem})$ where:
\be

  \item $L$ is a formal language, i.e, a set of
    expressions.\footnote{No distinction is made between how
      expressions are constructed in this definition as well as in
      subsequent definitions.  In particular, expressions constructed
      by binding variables are not treated in any special way.}

  \item $D_{\rm sem}$ is a nonempty domain (set) of \emph{semantic
    values}.

  \item $V_{\rm sem} : L \tarrow D_{\rm sem}$ is a total function,
    called a \emph{semantic valuation function}, that assigns each
    expression $e \in L$ a semantic value $V_{\rm sem}(e) \in D_{\rm
      sem}$. \hfill$\Box$

\ee  
\end{definition}

A syntax representation of a formal language is an assignment of
syntactic values to the expressions of the language:

\begin{definition}[Syntax Representation] \label{df:syn-rep} \em
Let $L$ be a formal language. A \emph{syntax representation} of $L$ is
a pair $R=(D_{\rm syn},V_{\rm syn})$ where:
\be

  \item $D_{\rm syn}$ is a nonempty domain (set) of \emph{syntactic
    values}.  Each member of $D_{\rm syn}$ represents a syntactic
    structure.

  \item $V_{\rm syn} : L \tarrow D_{\rm syn}$ is an injective, total
    function, called a \emph{syntactic valuation function}, that
    assigns each expression $e \in L$ a syntactic value $V_{\rm
      syn}(e) \in D_{\rm syn}$ such that $V_{\rm syn}(e)$ represents
    the syntactic structure of $e$. \hfill$\Box$

\ee 
\end{definition}

A syntax language for a syntax representation is a language of
expressions that denote syntactic values in the syntax representation:

\begin{definition}[Syntax Language] \label{df:syn-lang} \em
Let $R=(D_{\rm syn},V_{\rm syn})$ be a syntax representation of a
formal language $L_{\rm obj}$.  A \emph{syntax language} for $R$ is a pair
$(L_{\rm syn}, I)$ where:
\be

  \item $I = (L,D_{\rm sem},V_{\rm sem})$ in an interpreted language.

  \item $L_{\rm obj}\subseteq L$, $L_{\rm syn} \subseteq L$, and
    $D_{\rm syn} \subseteq D_{\rm sem}$.

  \item $V_{\rm sem}$ restricted to $L_{\rm syn}$ is a total function
    $V'_{\rm sem} : L_{\rm syn} \tarrow D_{\rm syn}$. \hfill$\Box$

\ee
\end{definition}

Finally, we are now ready to define a syntax framework:

\begin{definition}[Syntax Framework in an Interpreted Language] \label{df:syn-frame-lang} \em
\bsp Let $I=(L,D_{\rm sem},V_{\rm sem})$ be an interpreted language
and $L_{\rm obj}$ be a sublanguage of $L$.  A \emph{syntax framework}
for $(L_{\rm obj},I)$ is a tuple $F=(D_{\rm syn},V_{\rm syn}, L_{\rm
  syn}, Q, E)$ where: \esp

\be

  \item $R = (D_{\rm syn},V_{\rm syn})$ is a syntax representation of
    $L_{\rm obj}$.

  \item $(L_{\rm syn},I)$ is a syntax language for $R$.

  \item $Q : L_{\rm obj} \tarrow L_{\rm syn}$ is an injective, total
    function, called a \emph{quotation function}, such that:

    \textbf{Quotation Axiom.} For all $e \in L_{\rm obj}$, \[V_{\rm
      sem}(Q(e)) = V_{\rm syn}(e).\]

  \item $E : L_{\rm syn} \tarrow L_{\rm obj}$ is a (possibly partial)
    function, called an \emph{evaluation function}, such that:

    \textbf{Evaluation Axiom.} For all $e \in L_{\rm syn}$, \[V_{\rm
      sem}(E(e)) = V_{\rm sem}(V_{\rm syn}^{-1}(V_{\rm sem}(e)))\]
    whenever $E(e)$ is defined. \hfill$\Box$

\ee 
\end{definition}

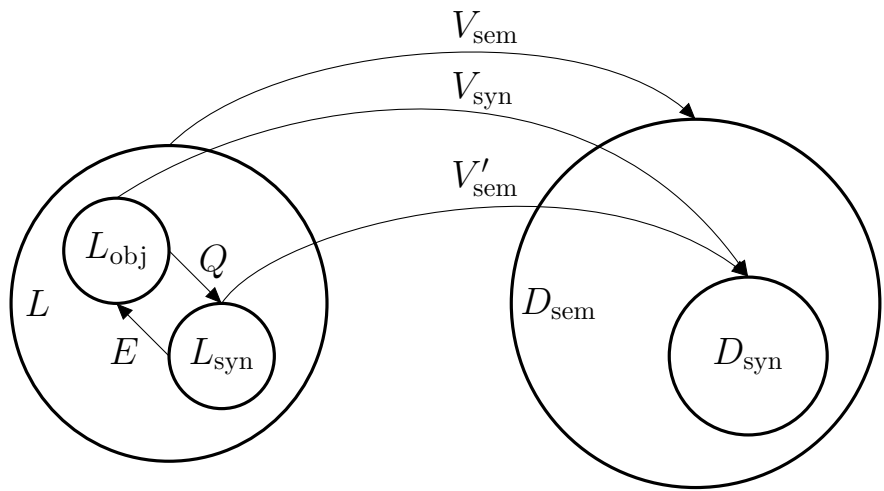
\begin{figure}
\center
\begin{tikzpicture}[scale=.70]
  \draw[very thick] (-3,0) circle (3);
    \draw (-5.5,0) node {\Large $L$};
  \draw[very thick] (-4,1) circle (1);
    \draw (-4,1) node {\Large $L_{\rm obj}$};
  \draw[very thick] (-2,-1) circle (1);
    \draw (-2,-1) node {\Large $L_{\rm syn}$};
  \draw[very thick] (7,0) circle (3.5);
    \draw (4.4,0) node {\Large $D_{\rm sem}$};
  \draw[very thick] (8,-1) circle (1.5);
    \draw (8,-1) node {\Large $D_{\rm syn}$};
  \draw[-triangle 45] (-3,3) .. controls (-1,5) and (5,5.5) .. (7,3.5);
    \draw[right] (2.2,5.25) node {\Large $V_{\rm sem}$};
  \draw[-triangle 45] (-4,2) .. controls (-1,4) and (5,5) .. (8,.5);
    \draw[right] (2.2,4.05) node {\Large $V_{\rm syn}$};
  \draw[-triangle 45] (-2,0) .. controls (-1,1.5) and (5,3) .. (8,.51);
    \draw[right] (2.2,2.4) node {\Large $V'_{\rm sem}$};
  \draw[-triangle 45] (-3,1) -- (-2,0);
    \draw[right] (-2.6,.8) node {\Large $Q$};
  \draw[-triangle 45] (-3,-1) -- (-4,0);
    \draw[right] (-4.3,-.9) node {\Large $E$};
\end{tikzpicture}
\caption{A Syntax Framework}  \label{fig:syn-frame}
\end{figure}

A syntax framework is depicted in Figure~\ref{fig:syn-frame}.  For $e
\in L_{\rm obj}$, $Q(e)$ is called the \emph{quotation} of $e$.
$Q(e)$ denotes a value in $D_{\rm syn}$ that represents the syntactic
structure of $e$.  For $e \in L_{\rm syn}$, $E(e)$ is called the
\emph{evaluation} of $e$.  If it is defined, $E(e)$ denotes the same
value in $D_{\rm sem}$ that the expression represented by the value of
$e$ denotes.  Since there will usually be different $e_1,e_2 \in
L_{\rm syn}$ that denote the same syntactic value, $E$ will usually
not be injective.  $Q$ and $E$ correspond to the quote and eval
operators in Lisp and other languages.

Common examples of syntax frameworks are based on representing the
syntax of expressions by G\"odel numbers, strings, and members of an
inductive type.  Programming languages that support metaprogramming
--- such as Lisp, F\#~\cite{FSharp13}, MetaML~\cite{TahaSheard00},
MetaOCaml~\cite{MetaOCaml11}, reFLect~\cite{GrundyEtAl06}, and
Template Haskell~\cite{SheardJones02} --- are instances of a syntax
framework if mutable variables are disallowed.
See~\cite{FarmerLarjani13} for these and other examples of syntax
frameworks.

The notion of a syntax framework can be easily lifted from an
interpreted language to an interpreted theory.  This is the version of
a syntax framework that we will use in this paper.

\begin{definition}[Model] \label{df:model} \em
Let $T = (L,\Gamma)$ be a theory.  A \emph{model} of $T$ is a pair $M
= (D^{M}_{\rm sem},V^{M}_{\rm sem})$ such that $D^{M}_{\rm sem}$ is a
nonempty set of semantic values that includes the truth values $\TRUE$
(true) and $\FALSE$ (false) and $V^{M}_{\rm sem} : L \tarrow
D^{M}_{\rm sem}$ is a total function such that, for all sentences $A
\in \Gamma$, $V^{M}_{\rm sem}(A) = \TRUE$. \hfill$\Box$
\end{definition}

\begin{definition}[Interpreted Theory] \label{df:interp-thy} \em
An \emph{interpreted theory} is a pair $I=(T,\sM)$ where $T$ is a
theory and $\sM$ is a set of models of $T$.  (If $T = (L,\Gamma)$,
$(L,D^{M}_{\rm sem},V^{M}_{\rm sem})$ is obviously an interpreted
language for each $M \in \sM$.) \hfill$\Box$
\end{definition}

\begin{definition}[Syntax Framework in an Interpreted Theory] \label{df:syn-frame-thy} \em
\hspace{2ex}\\ Let $I=(T,\sM)$ be an interpreted theory where $T =
(L,\Gamma)$ and $L_{\rm obj} \subseteq L$.  A \emph{syntax framework}
for $(L_{\rm obj},I)$ is a triple $F=(L_{\rm syn}, Q, E)$ where:

\be

  \item $L_{\rm syn} \subseteq L$.

  \item $Q : L_{\rm obj} \tarrow L_{\rm syn}$ is an injective, total
    function.

  \item $E : L_{\rm syn} \tarrow L_{\rm obj}$ is a (possibly partial)
    function.

  \item For all $M = (D^{M}_{\rm sem},V^{M}_{\rm sem}) \in \sM$,
    $F^M=(D^{M}_{\rm syn},V^{M}_{\rm syn}, L_{\rm syn}, Q, E)$ is a
    syntax framework for $(L_{\rm obj},(L,D^{M}_{\rm sem},V^{M}_{\rm
      sem}))$ where $D^{M}_{\rm syn}$ is the range of $V^{M}_{\rm sem}$
    restricted to $L_{\rm syn}$ and $V^{M}_{\rm syn} = V^{M}_{\rm sem}
    \circ Q$.

\ee 
\end{definition}

\bsp Let $I=(L,D,V)$ be an interpreted language, $L_{\rm obj}
\subseteq L$, and $F=(D_{\rm syn}, V_{\rm syn}, L_{\rm syn}, Q, E)$ be
a syntax framework for $(L_{\rm obj},I)$.  $F$ has \emph{built-in
  quotation} if there is an operator (which we will denote as
\mname{quote}) such that, for all $e \in L_{\rm obj}$, $Q(e)$ is the
syntactic result of applying the operator to $e$ (which we will denote
as $\mname{quote}(e)$).  $F$ has \emph{built-in evaluation} if there
is an operator (which we will denote as \mname{eval}) such that, for
all $e \in L_{\rm syn}$, $E(e)$ is the syntactic result of applying
the operator to $e$ (which we will denote as $\mname{eval}(e)$)
whenever $E(e)$ is defined.  There are similar definitions of built-in
quotation and evaluation for syntax frameworks in interpreted
theories.  \esp

A syntax framework $F$ for $(L_{\rm obj},I)$, where $I$ is either an
interpreted language or an interpreted theory, is \emph{replete} if
$L_{\rm obj} = L$ and $F$ has both built-in quotation and evaluation.
If $F$ is replete, it has the facility to reason about the syntax of
all of $L$ within $L$ itself. Examples of a replete syntax framework
are rare.  The programming language Lisp with a simplified semantics
is the best known example of a replete syntax
framework~\cite{FarmerLarjani13}.  T. \AE. Mogensen's
self-interpretation of lambda calculus~\cite{Mogensen94} and the
logic Chiron~\cite{Farmer07}, derived from classical NBG set theory,
are two other examples of replete syntax
frameworks~\cite{FarmerLarjani13}.

\section{Local Approach} \label{sec:local}

In order to formalize $\mname{PolyDiff}$ in $T_R$ we need the ability
to reason about the polynomials in $L_{\rm poly}$ as syntactic
structures (i.e., as syntax trees).  This can be achieved by
constructing a syntax framework for $(L_{\rm poly},I'_R)$ where $I'_R
= (T'_R,\sM')$ is an interpreted theory such that $T'_R$ is a
conservative extension of $T_R$.  Since we seek to reason about just
the syntax of $L_{\rm poly}$ instead of a larger language, we call
this the \emph{local approach}.

The construction of the syntax framework requires the following steps:

\be

\item Define in $T_R$ an inductive type whose members are the syntax
    trees of the polynomials in $L_{\rm poly}$.  The inductive type
    should include a new type symbol $\mathbb{S}$ and appropriate
    constants for constructing and deconstructing expressions of type
    $\mathbb{S}$.  Let $L_{\rm syn}$ be the set of expressions of type
    $\mathbb{S}$.  For example, if $x + 3$ is a polynomial in $L_{\rm
      poly}$, then an expression like
    $\mname{plus}(\mname{var}(\mname{s}_x),\mname{con}(\mname{s}_3))$
    could be the expression in $L_{\rm syn}$ that denotes the syntax
    tree of $x + 3$.  Next add an unspecified ``binary'' constant
    $O_{\rm pd}$ of type $\mathbb{S} \tarrow (\mathbb{S} \tarrow
    \mathbb{S})$ to $L_R$ (that is intended to represent
    $\mname{PolyDiff}$).  Let $T'_R = (L'_R,\Gamma'_R)$ be the
    resulting extension of $T_R$.  $T'_R$ is clearly a conservative
    extension of $T_R$.

  \item In the metatheory of $T'_R$ define an injective, total
    function $Q : L_{\rm poly} \tarrow L_{\rm syn}$ such that, for
    each polynomial $u \in L_{\rm poly}$, $Q(u)$ is an expression $e$
    that denotes the syntax tree of $u$.  For example, $Q(x + 3)$ could
    be
    $\mname{plus}(\mname{var}(\mname{s}_x),\mname{con}(\mname{s}_3))$.

  \item In the metatheory of $T'_R$ define a total mapping $E : L_{\rm
    syn} \tarrow L_{\rm poly}$ such that, for each expression $e \in
    L_{\rm syn}$, $E(e)$ is the polynomial whose syntax tree is
    denoted by $e$.  For example,
    $E(\mname{plus}(\mname{var}(\mname{s}_x),\mname{con}(\mname{s}_3)))$
    would be $x + 3$.

\ee
Let $(L_{\rm poly},I'_{R})$ where $I'_{R} = (T'_{R},\sM')$ and $\sM'$
is the set of standard models of $T'_{R}$ in simple type theory
(see~\cite{Farmer08}).  It is easy to check that $F = (L_{\rm
  syn},Q,E)$ is a syntax framework for $(L_{\rm poly},I'_R)$.  Notice
that $E$ is the left inverse of $Q$ and hence the \emph{law of
  disquotation} holds: For all $u \in L_{\rm poly}$, $E(Q(u)) = u$.

We are now ready to formalize $\mname{PolyDiff}$ in $T'_R$.  First, we
need to define an operator in $T'_{R}$ to represent
$\mname{PolyDiff}$.  We will use $O_{\rm pd}$ for this purpose.  We
write a sentence \mname{CompBehavior} \[\LambdaApp a,b \mcolon
\mathbb{S} \mdot \mname{is-var}(b) \ImpliesAlt B(a,b,O_{\rm
  pd}(a)(b))\] in $T'_R$ where, for all $u \in L_{\rm poly}$ and $x
\in L_{\rm var}$, \[B(Q(u),Q(x),O_{\rm pd}(Q(u))(Q(x)))\] holds iff
\[\mname{PolyDiff}(u,x) = E(O_{\rm pd}(Q(u))(Q(x))).\]  That is, we
specify the computational behavior of $O_{\rm pd}$ to be the same as
that of \mname{PolyDiff}.

Second, we need to prove that $O_{\rm pd}$ is mathematically correct.
We write the sentence \mname{MathMeaning} \[\mbox{for all\ }u \in
L_{\rm poly}, \mname{deriv}(\LambdaApp x \mcolon \mathbb{R} \mdot u) =
\LambdaApp x \mcolon \mathbb{R} \mdot E(O_{\rm pd}(Q(u))(Q(x)))\] in
the metatheory of $T'_R$ that says $O_{\rm pd}$ computes a syntactic
value that represents an expression that denotes the derivative of
$\LambdaApp x \mcolon \mathbb{R} \mdot u$ at $x$.  And then we prove
in $T'_R$ that \mname{MathMeaning} follows from \mname{CompBehavior}.
The proof requires showing that $E(O_{\rm pd}(Q(u))(Q(x)))$ equals
$\mname{deriv}((\LambdaApp x \mcolon \mathbb{R} \mdot u),x)$, which is
\[\lim_{h \tarrow 0}\frac{(\LambdaApp x \mcolon \mathbb{R} \mdot u)(x + h) -
  (\LambdaApp x \mcolon \mathbb{R} \mdot u)(x)}{h}.\] The details of
the proof are found in any good calculus textbook such
as~\cite{Spivak08}.

Third, we need to show how \mname{PolyDiff} can be used to compute the
derivative of a function $\LambdaApp x \mcolon \mathbb{R} \mdot u$ at
$x$ in $T'_R$.  There are two ways.  The first way is to simplify
$E(O_{\rm pd}(Q(u))(Q(x)))$ in \mname{MathMeaning} (e.g., by
beta-reduction).  The second way is to replace $E(O_{\rm
  pd}(Q(u))(Q(x)))$ in \mname{MathMeaning} with the result of applying
\mname{PolyDiff} to $u$ and $x$.  The first way requires that
\mname{PolyDiff} is implemented in $T'_R$ as $O_{\rm pd}$.  The second
way does not require that \mname{PolyDiff} is implemented in $T'_R$,
but only that its meaning is specified in $T'_R$.

The local approach is commonly used to reason about the syntax of
expressions in a formal theory.  It embodies a \emph{deep
  embedding}~\cite{BoultonEtAl93} of the object language (e.g.,
$L_{\rm poly}$) into the underlying formal language (e.g., $L_R$).
The local approach to reason about syntax can be employed in almost
any proof assistant in which it is possible to define an inductive
type (e.g., see
\cite{BoultonEtAl93,ContejeanEtAl07,WildmoserNipkow04}).

\bsp The local approach has both strengths and weaknesses.  These are the
strengths of the local approach: \esp

\be

  \item \emph{Indirect Reasoning about the syntax of $L_{\rm poly}$ in
    the Theory.}  In $T'_R$ using $L_{\rm syn}$, we can indirectly
    reason about the syntax of the polynomials in $L_{\rm poly}$.
    This thus enables us to specify the computational behavior of
    \mname{PolyDiff} via $O_{\rm pd}$.

  \item \emph{Direct Reasoning about the syntax of $L_{\rm poly}$ in
    the Metatheory.} In the metatheory of $T'_R$ using $L_{\rm syn}$,
    $Q$, and $E$, we can directly reason about the syntax of the
    polynomials in $L_{\rm poly}$.  In particular, using
    \mname{MathMeaning} and the formula \[\mbox{for all\ }u \in L_{\rm
      poly}, x \in L_{\rm var}, \mname{PolyDiff}(u,x) = E(O_{\rm
      pd}(Q(u))(Q(x))),\] we can specify the mathematical meaning of
    \mname{PolyDiff}.

\ee

And these are the weaknesses:

\be

  \item \emph{Syntax Problem.}  We cannot directly refer in $T'_R$ to
    the syntax of polynomials.  Also the variable $x$ is free in $x +
    3$ but not in $Q(x + 3) =
    \mname{plus}(\mname{var}(\mname{s}_x),\mname{con}(\mname{s}_3))$.
    As a result, $Q$ and $E$ cannot be defined in $T'_R$ and thus
    \mname{PolyDiff} cannot be fully formalized in $T'_R$.  In short,
    we can reason about syntax in $T'_R$ but not about the interplay
    of syntax and semantics in $T'_R$.

  \item \emph{Coverage Problem.} The syntax framework $F$ can only be
    used for reasoning about the syntax of polynomials.  It cannot be
    used for reasoning, for example, about rational expressions.  To
    do that a new syntax framework must be constructed.

  \item \emph{Extension Problem.} $L_{\rm poly}$, $L_{\rm syn}$, $Q$,
    and $E$ must be extended each time a new constant of type
    $\mathbb{R}$ is defined in $T'_R$ .

\ee
In summary, the local approach only gives us indirect access to the
syntax of polynomials and must be modified to cover new or enlarged
contexts.

If $L_{\rm obj}$ (which is $L_{\rm poly}$ in our example) does not
contain variables, then we can define $E$ to be a total operator in
the theory.  (If the theory is over a traditional logic, we will still
not be able to define $Q$ in the theory.)  This variant of the local
approach is used, for example, in the Agda reflection
mechanism~\cite{VanDerWalt12}.

\section{Global Approach} \label{sec:global}

The \emph{global approach} described in this section utilizes a
replete syntax framework.  Assume that we have modified $T_R$ and
simple type theory so that there is a replete syntax framework $F =
(L_{\rm syn},Q,E)$ for $(L_R,I_R)$ where $I_R = (T_R,\sM)$ and $\sM$
is the set of standard models of $T_R$ in the modified simple type
theory.  Let us also assume that $L_{\rm syn}$ is the set of
expressions of type $\mathbb{S}$ and $L_R$ includes a constant $O_{\rm
  pd}$ of type $\mathbb{S} \tarrow (\mathbb{S} \tarrow \mathbb{S})$.
By virtue of $F$ being replete, $F$ embodies a deep embedding of $L_R$
into itself.

As far as we know, no one has ever worked out the details of how to
modify simple type theory so that it admits built-in quotation and
evaluation for the full language of a theory.  However, we have shown
how NBG set theory can be modified to admit built-in quotation and
evaluation for its entire language~\cite{Farmer07}.  Simple type
theory can be modified in a similar way.  We plan to present a version
of simple type theory with a replete syntax framework in a future
paper.

We can formalize $\mname{PolyDiff}$ in $T_R$ as follows.  We will
write $\mname{quote}(e)$ and $\mname{eval}(e)$ as $\synbrack{e}$ and
$\sembrack{e}$, respectively.  First, we define the operator $O_{\rm
  pd}$ in $T_{R}$ to represent $\mname{PolyDiff}$.  We write a
sentence \mname{CompBehavior} \[\LambdaApp a,b \mcolon \mathbb{S}
\mdot \mname{is-poly}(a) \And \mname{is-var}(b) \ImpliesAlt
B(a,b,O_{\rm pd}(a)(b))\] in $T_R$ where, for all $u \in L_{\rm poly}$
and $x \in L_{\rm var}$, \[B(\synbrack{u},\synbrack{x},O_{\rm
  pd}(\synbrack{u})(\synbrack{x}))\] holds iff \[\mname{PolyDiff}(u,x)
= \sembrack{O_{\rm pd}(\synbrack{u})(\synbrack{x})}.\] That is, we
specify the computational behavior of $O_{\rm pd}$ to be the same as
that of \mname{PolyDiff}.

Second, we prove in $T_{R}$ that $O_{\rm pd}$ is mathematically
correct.  We write the sentence \mname{MathMeaning} \[\ForallApp a
\mcolon \mathbb{S} \mdot \mname{is-poly}(a) \ImpliesAlt
\mname{deriv}(\LambdaApp x \mcolon \mathbb{R} \mdot \sembrack{a}) =
\LambdaApp x \mcolon \mathbb{R} \mdot \sembrack{O_{\rm
    pd}(a)(\synbrack{x})}\] in $T_R$ that says $O_{\rm pd}$ computes a
syntactic value that represents an expression that denotes the
derivative of $\LambdaApp x \mcolon \mathbb{R} \mdot \sembrack{a}$
at $x$.  And then we prove in $T_R$ that \mname{MathMeaning} follows
from \mname{CompBehavior}.

Third, we use \mname{PolyDiff} to compute the derivative of a function
$\LambdaApp x \mcolon \mathbb{R} \mdot u$ at $x$ in $T_R$ in either
of the two ways described for the local approach.

The strengths of the global approach are:

\be

  \item \emph{Direct Reasoning about the syntax of polynomials in the
    Theory.}  In $T_R$ using $L_{\rm syn}$, \mname{quote}, and
    \mname{eval}, we can directly reason about the syntax of the
    expressions in $L_{\rm poly}$.  As a result, we can formalize
    \mname{PolyDiff} in $T_R$ as described in the Introduction.

  \item \emph{Direct Reasoning about the syntax of all expressions in
    the Theory.}  In $T_R$ using $L_{\rm syn}$, \mname{quote}, and
    \mname{eval}, we can directly reason about the syntax of the
    expressions in the entire language $L_R$.  As a result, the syntax
    framework $F$ can cover all current and future syntax reasoning
    needs.  Moreover, we can express such things as syntactic side
    conditions, formula schemas, and substitution for a variable
    directly in $T_R$ (see~\cite{Farmer07} for details).

\ee
In short, not only does the global approach enable us to formalize
\mname{PolyDiff} in $T_R$, it provides us with the facility to move
syntax-based reasoning from the metatheory of $T_R$ to $T_R$ itself.
This seems to be a wonderful result that solves the problem of
formalizing syntax-based mathematical algorithms.  Unfortunately, the
global approach has the following serious weaknesses that temper the
enthusiasm one might have for its strengths:

\be

  \item \emph{Evaluation Problem.}\\[1ex] Claim: \mname{eval} cannot
    be defined on all expressions in $L_R$.\\[1ex] Proof: Suppose
    \mname{eval} is indeed total.  $T_R$ is sufficiently expressive,
    in the sense of G\"odel's incomplete theorem, so apply the
    diagonalization lemma~\cite{Carnap34} to obtain a formula
    $\mname{LIAR}$ such that \[\mname{LIAR} =
    \synbrack{\Neg\sembrack{\mname{LIAR}}}.\]
    Then \[\sembrack{\mname{LIAR}} =
    \sembrack{\synbrack{\Neg\sembrack{\mname{LIAR}}}} =
    \Neg\sembrack{\mname{LIAR}},\] which is a
    contradiction. \hfill$\Box$\\[1ex] This means that the liar paradox
    limits the use of \mname{eval} and, in particular, the law of
    disquotation does not hold universally, i.e., there are
    expressions $e$ in $L_R$ such that $\sembrack{\synbrack{e}} \not=
    e$.

  \item \emph{Variable Problem.}  The variable $x$ is not free in the
    expression $\synbrack{x + 3}$ (or in any quotation).  However, $x$
    is free in $\sembrack{\synbrack{x + 3}}$ because
    $\sembrack{\synbrack{x + 3}} = x + 3$.  If the value of the
    variable $e$ is $\synbrack{x + 3}$, then both $e$ and $x$ are free
    in $\sembrack{e}$ because $\sembrack{e} = \sembrack{\synbrack{x +
        3}} = x + 3$.\\[1ex] This example shows that the notions of a
    free variable, substitution for a variable, etc.\ are
    significantly more complex when expressions contain \mname{eval}.

  \item \emph{Extension Problem.}  We can define $L_{\rm con}
    \subseteq L_{\rm syn}$ in $T_R$ as the language of expressions
    denoting the syntactic values of constants in $L_R$.\\[1ex] Claim:
    Assume the set of constants in $L$ is finite and
    $T'_R=(L'_R,\Gamma'_R)$ is an extension of $T_R$ such that there
    is a constant in $L'_R$ but not in $L_R$.  Then $T'_R$ is not a
    conservative extension of $T_R$.\\[1ex] Proof: Let
    $\set{c_1,\ldots,c_n}$ be the set of constants in $L$.  Then
    \[L_{\rm con} = \set{\synbrack{c_1},\ldots,\synbrack{c_n}}\] 
    is valid in $T_R$ but not in $T'_R$. \hfill$\Box$\\[1ex] This
    shows that in the global approach the development of a theory via
    definitions requires that the notion of a conservative extension
    be weakened.

  \item \emph{Interpretation Problem.}\\[1ex] Let $T = (L,\Gamma)$ and
    $T' = (L',\Gamma')$ in be two theories in a simple type theory
    that has been modified to admit built-in quotation and evaluation
    for the entire language of a theory.\\[1ex] Claim: Let $\Phi$ be
    an interpretation of $T$ in $T'$ such that $\Phi$ is a
    homomorphism with respect to the logical operators of the
    underlying logic.  Then $\Phi$ must be injective on the constants
    of $L$.\\[1ex] Proof: Assume that $\Phi$ is not injective on
    constants.  Then there are two different constants $a,b$ such that
    $\Phi(a) = \Phi(b)$.  $\synbrack{a} \not= \synbrack{b}$ is valid
    in $T$.  Hence \[\Phi(\synbrack{a} \not= \synbrack{b}) =
    (\synbrack{\Phi(a)} \not= \synbrack{\Phi(b)})\] since $\Phi$ is a
    homomorphism, and the latter inequality must be valid in $T'$
    since $\Phi$ is an interpretation (which maps valid formulas of
    $T$ to valid formulas of $T'$).  However, our hypothesis $\Phi(a)
    = \Phi(b)$ implies $\synbrack{\Phi(a)} = \synbrack{\Phi(b)}$,
    which is a contradiction. \hfill$\Box$\\[1ex] This shows that the
    use of interpretations is more cumbersome in a logic that admits
    quotation than one that does not.

\ee


\section{Conclusion} \label{sec:conclusion}

Syntax-based mathematical algorithms are employed throughout
mathematics and are one of the main offerings of computer algebra
systems.  They are difficult, however, to formalize since they
manipulate the syntactic structure of expressions in mathematically
meaningful ways.  We have presented two approaches to formalizing
syntax-based mathematical algorithms in a formal theory, one called
the \emph{local approach} and the other the \emph{global approach}.
Both are based on the notion of a \emph{syntax framework} which
provides a foundation for integrating reasoning about the syntax of
expressions with reasoning about what the expressions mean.  Syntax
frameworks include a syntax representation, a syntax language for
reasoning about the representation, and quotation and evaluation
mechanisms.  Common syntax reasoning systems are instances of a syntax
framework.

The local approach and close variants are commonly used for
formalizing syntax-based mathematical algorithms.  Its major strength
is that it provides the means to formally reason about the syntactic
structure of expressions, while its major weakness is that the
mathematical meaning of a syntax-based mathematical algorithm cannot
be expressed in the formal theory.  Another weakness is that an
application of the local approach cannot be easily extended to cover
new or enlarged contexts.

The global approach enables one to reason in a formal theory $T$
directly about the syntactic structure of the expressions in $T$ as
well as about the interplay of syntax and semantics in $T$.  As a
result, it is possible to fully formalize syntax-based algorithms like
\mname{PolyDiff} and move syntax-based reasoning, like the use of
syntactic side conditions, from the metatheory of $T$ to $T$ itself.
Unfortunately, these highly desirable results come with a high cost:
Significant change must be made to the underlying logic as illustrated
by the Evaluation, Variable, Extension, and Interpretation Problems
given in the previous section.  

One of the main goals of the MathScheme
project~\cite{CaretteFarmer11}, led by J. Carette and the author, is
to see if the global approach can be used as a basis to integrate
axiomatic and algorithmic mathematics.  The logic
Chiron~\cite{Farmer07} demonstrates that it is possible to modify a
traditional logic to support the global approach.  Although we have
begun an implementation of Chiron, it remains an open question whether
a logic modified in this way can be effectively implemented.  As part
of the MathScheme project, we are now pursuing this problem as well as
developing the techniques needed to employ the global approach.

\bigskip

\section*{Acknowledgments.}  

The author would like to thank Jacques Carette and Pouya Larjani for
many fruitful discussions on ideas related to this paper.  The author
is also grateful to the referees for their comments and careful review
of the paper.

\bibliography{$HOME/research/lib/imps} \bibliographystyle{plain}

\end{document}

%% file: llncs-def.tex
\newcommand{\be}{\begin{enumerate}}
\newcommand{\ee}{\end{enumerate}}
\newcommand{\bi}{\begin{itemize}}
\newcommand{\ei}{\end{itemize}}
\newcommand{\bc}{\begin{center}}
\newcommand{\ec}{\end{center}}
\newcommand{\bsp}{\begin{sloppypar}}
\newcommand{\esp}{\end{sloppypar}}

\newcommand{\sM}{\mbox{$\cal M$}}

\renewcommand{\phi}{\varphi}

\newcommand{\set}[1]{{\{ #1 \}}}

\newcommand{\sembrack}[1]{\llbracket#1\rrbracket}
\newcommand{\synbrack}[1]{\ulcorner#1\urcorner}

\newcommand{\mname}[1]{\mbox{\sf #1}}
\newcommand{\mcolon}{\mathrel:}
\newcommand{\mdot}{\mathrel.}

\newcommand{\tarrow}{\rightarrow}

\newcommand{\Neg}{\neg} 
\newcommand{\And}{\wedge}

\newcommand{\ImpliesAlt}{\Rightarrow}

\newcommand{\ForallApp}{\forall\,}

\newcommand{\TRUE}{\mbox{{\sc t}}}
\newcommand{\FALSE}{\mbox{{\sc f}}}
\newcommand{\LambdaApp}{\lambda\,}

\newcommand{\funapp}{\mathrel@}

%% file: quote-eval.bbl
\begin{thebibliography}{10}

\bibitem{BoultonEtAl93}
R.~Boulton, A.~Gordon, M.~Gordon, J.~Harrison, J.~Herbert, and J.~Van Tassel.
\newblock Experience with embedding hardware description languages in {HOL}.
\newblock In V.~Stavridou, T.~F. Melham, and R.~T. Boute, editors, {\em
  Proceedings of the {IFIP} {TC10/WG 10.2} International Conference on Theorem
  Provers in Circuit Design: Theory, Practice and Experience}, volume A-10 of
  {\em IFIP Transactions A: Computer Science and Technology}, pages 129--156.
  North-Holland, 1993.

\bibitem{CaretteFarmer11}
J.~Carette and W.~M. Farmer.
\newblock Mathscheme: Project description.
\newblock In J.~H. Davenport, W.~M. Farmer, F.~Rabe, and J.~Urban, editors,
  {\em Intelligent Computer Mathematics}, volume 6824 of {\em Lecture Notes in
  Computer Science}, pages 287--288. Springer-Verlag, 2011.

\bibitem{Carnap34}
R.~Carnap.
\newblock {\em Die {L}ogische {S}yntax der {S}prache}.
\newblock Springer-Verlag, 1934.

\bibitem{ContejeanEtAl07}
E.~Contejean, P.~Courtieu, J.~Forest, O.~Pons, and X.~Urbain.
\newblock Certification of automated termination proofs.
\newblock In {\em Frontiers of Combining Systems}, volume 4720 of {\em Lecture
  Notes in Computer Science}, pages 148--162. Springer, 2007.

\bibitem{DemersMalenfant95}
F.-N. Demers and J.~Malenfant.
\newblock Reflection in logic, functional and object-oriented programming: A
  short comparative study.
\newblock In {\em IJCAI '95 Workshop on Reflection and Metalevel Architectures
  and their Applications in AI}, pages 29--38, 1995.

\bibitem{Farmer07b}
W.~M. Farmer.
\newblock Biform theories in {Chiron}.
\newblock In M.~Kauers, M.~Kerber, R.~R. Miner, and W.~Windsteiger, editors,
  {\em Towards Mechanized Mathematical Assistants}, volume 4573 of {\em Lecture
  Notes in Computer Science}, pages 66--79. Springer-Verlag, 2007.

\bibitem{Farmer07}
W.~M. Farmer.
\newblock Chiron: {A} set theory with types, undefinedness, quotation, and
  evaluation.
\newblock SQRL Report No.~38, McMaster University, 2007.
\newblock Revised 2012. Available at
  \url{http://imps.mcmaster.ca/doc/chiron-tr.pdf}.

\bibitem{Farmer08}
W.~M. Farmer.
\newblock The seven virtues of simple type theory.
\newblock {\em Journal of Applied Logic}, 6:267--286, 2008.

\bibitem{FarmerLarjani13}
W.~M. Farmer and P.~Larjani.
\newblock Frameworks for reasoning about syntax that utilize quotation and
  evaluation.
\newblock McSCert Report No.~9, McMaster University, 2013.
\newblock Available at \url{http://imps.mcmaster.ca/doc/syntax.pdf}.

\bibitem{FSharp13}
The F\#~Software Foundation.

\bibitem{GrundyEtAl06}
J.~Grundy, T.~Melham, and J.~O'Leary.
\newblock A reflective functional language for hardware design and theorem
  proving.
\newblock {\em Journal of Functional Programming}, 16, 2006.

\bibitem{Harrison95}
J.~Harrison.
\newblock Metatheory and reflection in theorem proving: A survey and critique.
\newblock Technical Report CRC-053, SRI Cambridge, 1995.
\newblock Available at
  \url{http://www.cl.cam.ac.uk/~jrh13/papers/reflect.ps.gz}.

\bibitem{Koellner09}
P.~Koellner.
\newblock On reflection principles.
\newblock {\em Annals of Pure and Applied Logic}, 157:206--219, 2009.

\bibitem{Mogensen94}
T.~\AE. Mogensen.
\newblock Efficient self-interpretation in lambda calculus.
\newblock {\em Journal of Functional Programming}, 2:345--364, 1994.

\bibitem{MetaOCaml11}
{Rice University Programming Languages Team}.
\newblock Metaocaml: A compiled, type-safe, multi-stage programming language.
\newblock \url{http://www.metaocaml.org/}, 2011.

\bibitem{SheardJones02}
T.~Sheard and S.~P. Jones.
\newblock Template meta-programming for {Haskell}.
\newblock {\em ACM SIGPLAN Notices}, 37:60--75, 2002.

\bibitem{Spivak08}
M.~Spivak.
\newblock {\em Calculus}.
\newblock Publish or Perish, fourth edition, 2008.

\bibitem{TahaSheard00}
W.~Taha and T.~Sheard.
\newblock Meta{ML} and multi-stage programming with explicit annotations.
\newblock {\em Theoretical Computer Science}, 248:211--242, 2000.

\bibitem{VanDerWalt12}
P.~van~der Walt.
\newblock Reflection in {Agda}.
\newblock Master's thesis, Universiteit Utrecht, 2012.

\bibitem{WildmoserNipkow04}
M.~Wildmoser and T.~Nipkow.
\newblock Certifying machine code safety: Shallow versus deep embedding.
\newblock In K.~Slind, A.~Bunker, and G.~Gopalakrishnan, editors, {\em TPHOLs},
  volume 3223 of {\em Lecture Notes in Computer Science}, pages 305--320.
  Springer-Verlag, 2004.

\end{thebibliography}
